\documentclass[12pt]{article}

\usepackage{newtxtext,newtxmath}

\usepackage{graphicx}

\usepackage[letterpaper,margin=1in]{geometry}

\renewenvironment{abstract}
	{\quotation}
	{\endquotation}

\date{}

\makeatletter
\renewcommand{\fnum@figure}{\textbf{Figure \thefigure}}
\renewcommand{\fnum@table}{\textbf{Table \thetable}}
\makeatother

\usepackage{scicite}

\usepackage{url}

\makeatletter
\newcommand{\addbar@}[3]{%
  \makebox[0pt][l]{%
    \raisebox{#1}[0pt][0pt]{%
      \kern#2
      \rule[#3]{0.365em}{#3}%
    }%
  }%
}
\DeclareRobustCommand{\qbar}{\text{\addbar@{-0.35ex}{0.09em}{0.083ex}}q}
\makeatother

\makeatletter
\newcommand{\llangle}{\mathopen{\langle\kern-0.2em\langle}}
\newcommand{\rrangle}{\mathclose{\rangle\kern-0.2em\rangle}}
\makeatother

\makeatletter
\newcommand{\llangleBig}{\mathopen{\Big\langle\kern-0.22em\Big\langle}}
\newcommand{\rrangleBig}{\mathclose{\Big\rangle\kern-0.22em\Big\rangle}}
\makeatother

\def\scititle{
	The limits of knowledge in classical physics resemble the quantum uncertainty relation
}
\title{\bfseries \boldmath \scititle}

\author{
	David Theurel$^{1\ast}$\and
	\small$^{1}$Helen Wills Neuroscience Institute, University of California, Berkeley 94720, USA.\and
	\small$^\ast$Email: theurel@alum.mit.edu
}

\begin{document} 

\maketitle

\begin{abstract} \bfseries \boldmath
Building upon a recent analysis of the measurement process in Hamiltonian mechanics, this article investigates the Bayesian epistemology of classical physics---the landscape of accessible probability distributions over phase space. I prove a thermodynamic limitation on the information that can be obtained about a classical system by means of observations: A direct analogue of the Robertson-Schr\"odinger quantum uncertainty relation controls the acquisition of information at the classical microscale. Central to this theorem is the notion of the \emph{quality} of a measuring probe; a temperature-dependent strictly positive quantity that serves as a figure of merit of the probe, and that plays the role of $1/\hbar$ in the classical uncertainty relation. This study sets the stage for a new area of research into resource theories of classical measurement, in which high-quality measurements and high-information states of knowledge are the limited resources.
\end{abstract}

\noindent
Most scientists experience the world in the first person (presumably). Yet, in an act of transcendence, most scientific theories adopt the ``view from nowhere''~\cite{nagel1989view}; they describe the world objectively, detached from any particular viewpoint. Theories with broad scope, such as fundamental theories of physics~\cite{fn0}, still hope to explain the structural features of first-person experience (if not its subjective character), by recognizing the subject as just another object, and treating observations and measurements as physical interactions between objects. As it regards the acquisition of information and knowledge, this reductive program might be summarized by the slogans ``Bit from it,'' or ``Epistemology from ontology.'' While such a program has a long tradition in physics [notably in thermodynamics~\cite{szilard1929entropieverminderung, parrondo2015thermodynamics} and quantum mechanics~\cite{von2018mathematical,nielsen2002quantum}], it has, surprisingly, not yet been carried out for one lauded fundamental physical theory: classical Hamiltonian mechanics~\cite{fn1}. The present article constitutes, to my knowledge, the second serious step in this direction. I build upon a recent analysis of the measurement process in Hamiltonian mechanics~\cite{theurel2024incompatible}, to prove a thermodynamic limitation on the information that can be obtained about a classical system by means of observations.

In Ref.~\cite{theurel2024incompatible} I examined classical measurement as a process involving the joint evolution of an object-system and a finite-temperature measuring probe. The analysis revealed a positive parameter, $1/\qbar$, that serves as a figure of merit of even the most idealized probe. I call this parameter the \emph{quality} of the probe. Inverse quality, $\qbar$ (``q-bar''), has physical dimensions of action and is, loosely speaking, the ``symplectic width'' of the probe's ready-state probability distribution over phase space. In an ideal preparation limited only by thermal noise, it is given by
\begin{equation}\label{eq_qbar}
	\qbar=\frac{k_\text{B}T}{\pi f},
\end{equation}
where $k_\text{B}T$ is the probe's temperature times Boltzmann's constant, and $f$ is the natural frequency of oscillation of the probe in its trapped ready state. The infimum value $\qbar=0$ requires absolute-zero temperature, and is therefore unreachable, by the third law of thermodynamics. For temperatures and frequencies in the broad ranges of $10^{-10}$--$10^3~\text{K}$ and $10^0$--$10^{17}~\text{Hz}$, which I expect cover most laboratory measurement preparations, Eq.~\ref{eq_qbar} yields values of $\qbar$ in the range $10^{-17}$--$10^{13}~\hbar$, in units of Planck's constant $\hbar\sim 1\times 10^{-34}$~J\,s. On the scale of every-day human experience these are all very small quantities of action---a circumstance which may partly account for the curious fact that a theory of classical measurement was not developed many decades before Ref.~\cite{theurel2024incompatible}. [See also~\cite{fn2}.] At the same time, said range of action suggests that the classical/thermal effects to which I am referring (quantified by $\qbar$) can in some cases be of comparable magnitude or dominate over quantum effects (quantified by $\hbar$). And indeed, in gravitational wave detection, for example, the joint importance of quantum and thermal noises is well known, and their parallel forms have recently been emphasized~\cite{whittle2023unification}. Pursuing a theory of classical measurement, in addition to being interesting in its own right, thus has the potential to improve our understanding of, and intuition for, real-world high-precision measurements. In particular, profound implications seem likely for the field of information thermodynamics~\cite{parrondo2015thermodynamics, fn1}.

The present article is addressed at the following question which has become open to examination with the new theory of measurement. For a hypothetical agent living in a world governed by classical Hamiltonian mechanics, how well can they come to know the state of the world by means of observations and measurements? In philosophy jargon: What is the (Bayesian) \emph{epistemology} of the \emph{ontology} given by classical Hamiltonian mechanics? It will be explained how such philosophical question can be made into a well-posed mathematical problem in a resource theory of classical measurement. And three preliminary results towards this problem's solution will be presented: (i)~an example of an explicitly solved measurement protocol, clearly exhibiting the boundary between reachable and unreachable states of knowledge; (ii)~a general theorem on the impossibility of gaining information beyond a certain bound; and (iii)~a corollary in the special case of systems with a single degree of freedom, for the impossibility of reaching certain states of knowledge. In all three cases, the key criterion (between reachable and unreachable, possible and impossible) will be whether or not the inequality holds in a classical analogue of the quantum uncertainty relation. These results suggest that said classical uncertainty relation (CUR) may turn out to be a general epistemic restriction; i.e., that it may be altogether impossible to reach states of knowledge that violate the CUR. Such a strong conjecture remains tentative at this moment, but these preliminary results already reveal a striking resemblance between the epistemology of classical physics and the quantum formalism, that deserves further attention.

Certain mathematical identities in harmonic analysis have been referred to as ``uncertainty relations''~\cite{folland1997uncertainty, havin2012uncertainty}. Unlike them, the present CUR is not an identity, but rather is a condition on the possibility of informative measurements, and may turn out to be a characterization of those states of knowledge that can be reached by means of measurements. In my understanding, the present CUR is also not directly related to the various thermodynamic uncertainty relations (TURs)~\cite{barato2015thermodynamic, horowitz2020thermodynamic}; the CUR being more closely analogous, in both form and meaning, to the uncertainty relation of quantum mechanics. (But see the Discussion for a possible application of the TURs to the present research program.) Finally, this article is not an attempt to ``derive'' quantum mechanics from classical mechanics, nor to challenge the validity of quantum mechanics, nor anything of that sort---this article is not \emph{about} quantum mechanics~\cite{fn3}.

\subsection*{Background: The canonical model of classical measurement}
In this section I summarize the main concepts and results regarding classical measurement, as developed in Ref.~\cite{theurel2024incompatible}. 

Consider a Hamiltonian system with $2n$-dimensional phase space $M=\{(\boldsymbol q,\boldsymbol p)\}=\{\boldsymbol z\}$ and Hamiltonian $H(\boldsymbol z)$. The \emph{observables} of this classical system are represented by real-valued single-valued smooth functions defined globally over phase space. The simplest model for measuring an observable $A(\boldsymbol z)$ involves an interaction of the \emph{canonical} (or product, or von Neumann) form,
\begin{equation}\label{eq_canonicalModel}
	H_\text{int}\propto A(\boldsymbol z)P,
\end{equation}
where $P$ is the \emph{probe} degree of freedom belonging to the measuring apparatus. I will refer to this as a \emph{canonical measurement}, by direct analogy with canonical quantum measurements~\cite{ozawa1993canonical}. If a canonical measurement of $A$ is done with a probe of quality $1/\qbar$, then the measurement \emph{imprecision}, $\epsilon_A$, and the magnitude of the \emph{disturbance} inflicted upon the system, $\eta_A$, have been shown to obey a Heisenberg-like precision-disturbance relation~\cite{fn4,fn5}:
\begin{equation}\label{eq_precDistRel}
	\epsilon_A\eta_A\geq\frac{\qbar}{2}.
\end{equation}
An \emph{infinitesimal measurement} is one taking place over a vanishing time step $dt\to0$, having infinitesimal precision (infinite imprecision) $\epsilon_A\propto1/\sqrt{dt}$ and disturbing the system only infinitesimally $\eta_A\propto\sqrt{dt}$. A \emph{continuous measurement} is a succession of infinitesimal measurements taking place over a period of time. Such a measurement produces a continuous stream of noisy measurement \emph{outcomes} $A^*(t)$, according to~\cite{fn6}
\begin{equation}
	A^*=\langle A\rangle+\frac{1}{\sqrt{8k}}\frac{dW}{dt}.
\end{equation}
Here $k\geq0$ is the measurement \emph{strength} (with physical dimensions of $[k]=[A]^{-2}\cdot\text{time}^{-1}$), the noise $W(t)=\int_0^tdW$ is a standard Wiener process, and angled brackets denote the phase-space average:
\begin{equation}
	\langle A\rangle=\int_{M} d^{2n}\boldsymbol z\rho(\boldsymbol z)A(\boldsymbol z),
\end{equation}
where $\rho(\boldsymbol z)$ is a probability distribution function (PDF) over phase space, quantifying the \emph{state of knowledge} or \emph{epistemic state} of an agent reading the measurement outcome~\cite{fn7}. The measurement is said to have \emph{efficiency} $\nu$ if the outcomes are successfully registered by the apparatus a fraction $\nu\in[0,1]$ of the time, and \emph{discarded} (missed) the rest of the time. The effect of such a continuous canonical measurement on the state of knowledge of a rational agent was shown to be given by the It\^o stochastic partial differential equation (PDE)
\begin{align}
	\frac{\partial\rho}{\partial t}&=
	\underbrace{\{H,\rho\}}_{\substack{\text{Liouville dynamics}\\ \text{Info.~preserved}}}
	+\underbrace{k\qbar^2\{A,\{A,\rho\}\}}_{\substack{\text{Disturbance}\\ \text{Diffusion along flow $\Phi^A_\tau$}\\ \text{Info.~of compatible observs.~preserved}\\ \text{Info.~of incompatible observs.~lost}}}\notag\\
	&\quad+\underbrace{\sqrt{8\nu k}(A-\langle A\rangle)\rho\,\frac{dW}{dt}}_{\substack{\text{Bayesian update} \\\text{Collapse towards measurement outcome}\\ \text{Nonlinear, nonlocal \& stochastic}\\{\llangle\Delta\text{info}\rrangle\geq0}}},\label{eq_masterEquation}
\end{align}
where curly brackets denote the Poisson bracket:
\begin{equation}
	\{A,B\}=\sum_{j=1}^n\left(\frac{\partial A}{\partial q_j}\frac{\partial B}{\partial p_j}-\frac{\partial A}{\partial p_j}\frac{\partial B}{\partial q_j}\right).
\end{equation}

Let the covariance matrix of a pair of observables $A,B$ be denoted by
\begin{align}
	\boldsymbol\Sigma_{AB}&=\left\langle
	\begin{pmatrix}
		A-\langle A\rangle\\
		B-\langle B\rangle
	\end{pmatrix}
	\begin{pmatrix}
		A-\langle A\rangle	&B-\langle B\rangle
	\end{pmatrix}
	\right\rangle\label{eq_covarianceMatrix1}\\
	&=
	\begin{pmatrix}
		\sigma_A^2	&\text{cov}(A,B)\\
		\text{cov}(A,B)	&\sigma_B^2
	\end{pmatrix},
\end{align}
and let the Gibbs-Shannon information of $\rho(\boldsymbol z)$ be denoted by
\begin{equation}\label{eq_GibbsShannon}
	I=\langle\log\rho\rangle.
\end{equation}
Under the dynamics of Eq.~\ref{eq_masterEquation}, information has been shown to evolve according to the stochastic differential equation (SDE)
\begin{align}
	\dot I&=4\nu k\sigma_A^2-k\qbar^2\sigma_{\{A,\log\rho\}}^2+\sqrt{8\nu k}\,\text{cov}(A,\log\rho) \frac{dW}{dt}.\label{eq_antiHtheorem1}
\end{align}
Letting double angled brackets denote averaging over the noise $dW$ [i.e., averaging over the upcoming measurement outcome $A^*(t)$ conditional on the measurement record $\{A^*(t')\,|\,t'< t\}$]:
\begin{equation}\label{eq_antiHtheorem2}
	\llangle\dot I\rrangle=\underbrace{4\nu k \sigma_A^2}_{\substack{\text{Bayesian update}\\ \llangle \Delta I\rrangle\geq0}}\underbrace{-\ \ k\qbar^2\sigma_{\{A,\log\rho\}}^2}_{\substack{\text{Disturbance}\\ \Delta I\leq0}}.
\end{equation}
Notice that it is possible for a measurement to yield a net loss of information about the system ($\dot I<0$), even in expectation ($\llangle\dot I\rrangle<0$), even for an ideal and perfectly efficient measurement ($\nu=1$). This peculiar situation happens whenever the gain of information about $A$ is overcome by the disturbance, which causes a loss of information regarding observables not in involution with~$A$.

\subsection*{Problem statement}

\subsubsection*{Classical measurement as a resource theory}

Building upon the results just reviewed, the theory of classical measurement may begin a natural next phase of its development as a resource theory of states and operations~\cite{coecke2016mathematical}. In this framework the states are epistemic states and the operations are Liouville dynamics and canonical measurements, as in Eq.~\ref{eq_masterEquation}. Many different such resource theories are possible, depending on which sets of states and operations are deemed as free and which as costly resources, perhaps tailored to particular experimental situations. For concreteness, in this paper I will adhere to the following ``coarsest'' resource theory, although the results derived will not be particularly sensitive to this choice. The state of complete ignorance (the uniform PDF over phase space) and all of Liouville dynamics are to be treated as free resources, while all other states and all measurements are costly resources. The central question is to determine which epistemic states can be reached---and which cannot---starting from the state of complete ignorance, by making free use of Liouville dynamics and performing measurements of quality no greater than some maximum $1/\qbar$. This is a well-posed mathematical problem~\cite{fn8}. And while I have been unable to arrive at a complete solution at this time, I will present here a number of partial results.

\subsubsection*{The classical uncertainty relation as a resource-theoretic conjecture}

The precision-disturbance relation~\ref{eq_precDistRel} is analogous to the Heisenberg precision-disturbance relation of quantum mechanics~\cite{fujikawa2012universally, busch2013proof, dressel2014certainty}, with $\qbar$ playing the role of $\hbar$. In quantum theory there is, additionally, the well known Heisenberg \emph{uncertainty} relation, which speaks to the limits of what can be known about the state of a quantum system~\cite{fn9}. For any two quantum observables $\hat A,\hat B$ it reads, in the Robertson-Schr\"odinger form~\cite{schrodinger1930heisenbergschen, steiger2010quantum, fn10}:
\begin{equation}\label{eq_QUR}
	\sigma_{\hat A}^2\sigma_{\hat B}^2-\text{cov}(\hat A,\hat B)^2\geq\left\langle\frac{1}{2i}[\hat A,\hat B]\right\rangle^2,
\end{equation}
where square brackets denote the quantum commutator. The quantum-classical parallel suggests that an analogous epistemic limitation might exist in the theory of classical measurement. Minding Dirac's recipe of canonical quantization~\cite{dirac1981principles}
\begin{equation}
	[\hat A,\hat B]\leftrightarrow i\hbar\{A,B\},
\end{equation}
the direct classical analogue of Eq.~\ref{eq_QUR}, with $\qbar$ playing the role of $\hbar$, would be
\begin{equation}\label{eq_CUR0}
	\sigma_A^2\sigma_B^2-\text{cov}(A,B)^2\geq\left\langle\frac{\qbar}{2}\{A,B\}\right\rangle^2.
\end{equation}
This is the CUR. But notice the sense in which such a candidate inequality must be understood. Unlike in the quantum formalism, there is nothing in the classical formalism that rules out the possibility of \emph{starting} with a state of knowledge that violates the CUR, for example
\begin{equation}
	\rho(\boldsymbol z)=\delta^{2n}(\boldsymbol z-\boldsymbol z_0).
\end{equation}
So Eq.~\ref{eq_CUR0} is obviously not valid as an identity. Rather, as anticipated above, the question is a resource-theoretic one: whether it is at all possible for a rational agent to \emph{arrive} at an epistemic state violating the CUR, starting from a state of ignorance, by means of measurements with quality no greater than $1/\qbar$.

\subsection*{Results}

\subsubsection*{Illustration of the uncertainty relation in an analytically solvable measurement protocol}

\begin{figure}
	\centering
	\includegraphics[width=0.6\linewidth]{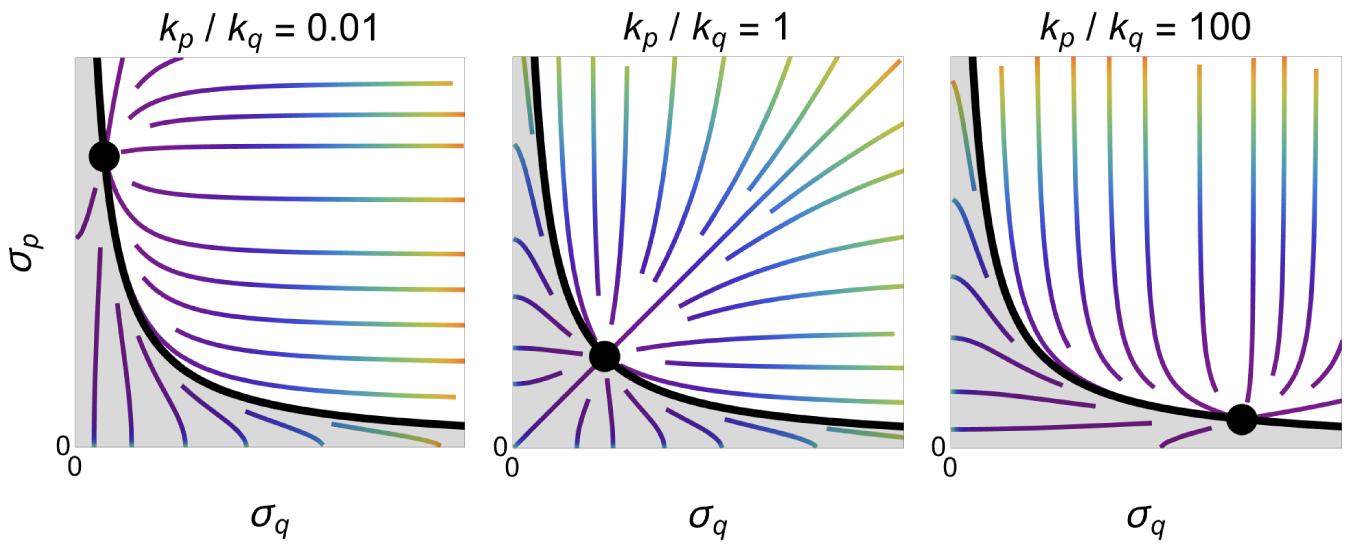} 
	\caption{\textbf{Reachable and unreachable states of knowledge in an analytically solvable measurement protocol.} (Color online.) Streamline plots of $(\sigma_{q}(t),\sigma_{p}(t))$ as given by Eqs.~\ref{eq_solSigma}, with warmer colors indicating higher stream speeds. Three panels show different choices of the ratio $k_p/k_q$. For any choice of $(k_q,k_p)$, all trajectories asymptote at the same terminal state (black dot). For all choices of $(k_q,k_p)$, terminal states lie along the hyperbola $\sigma_q\sigma_p=\qbar/2$ (black curve). Trajectories above (resp.~below) the hyperbola correspond to measurements that yield net gains (resp.~net losses) of information at each moment in time. Because trajectories do not cross the hyperbola, epistemic states violating Eq.~\ref{eq_classicalUncertaintyRel1} (shaded region) are unreachable starting from states of low information.}
	\label{fig_exact_solution_for_Gaussian_PDF}
\end{figure}
A special case of this question was answered in Ref.~\cite{theurel2024incompatible}. Working in the 2-dimensional phase space $\mathbb R^2=\{(q,p)\}$, Eq.~\ref{eq_masterEquation} was solved analytically for all future time starting from a simple class of initial PDFs which include the uniform distribution as a special case---namely, uncorrelated Gaussians in $q,p$---, for a simple class of measurement protocols: simultaneous~\cite{fn11} continuous canonical measurement of $q$ and $p$, at fixed quality $1/\qbar$, fixed respective strengths $k_q,k_p$, and perfect efficiencies $\nu_q=\nu_p=1$, in the absence of internal dynamics ($H\equiv0$). It was found that under such a protocol $\rho(q,p)$ remains Gaussian and $q,p$ remain uncorrelated for all time:
\begin{equation}\label{eq_explicitSol1}
	\rho(q,p,t)=\frac{1}{2\pi\sigma_{q}\sigma_{p}}\exp\left\{-\frac{(q-\mu_q)^2}{2\sigma_{q}^2}-\frac{(p-\mu_p)^2}{2\sigma_{p}^2}\right\},
\end{equation}
where the means are stochastic functions of time evolving as:
\begin{subequations}
\begin{align}
	d\mu_q&=\sqrt{8k_{q}}\sigma_{q}(t)^2dW_{q},\label{eq_explicitSol2}\\
	d\mu_p&=\sqrt{8k_{p}}\sigma_{p}(t)^2dW_{p},\label{eq_explicitSol3}
\end{align}
\end{subequations}
with $W_q, W_p$ independent Wiener processes, while the variances are the deterministic functions of time:
\begin{subequations}\label{eq_solSigma}
\begin{align}
	\sigma_{q}(t)^2&=\frac{\qbar}{2}\sqrt{\frac{k_p}{k_q}}\left[\coth\left\{4\qbar\sqrt{k_qk_p}(t-t_q)\right\}\right]^{l_q},\label{eq_solSigmaq}\\
	\sigma_{p}(t)^2&=\frac{\qbar}{2}\sqrt{\frac{k_q}{k_p}}\left[\coth\left\{4\qbar\sqrt{k_qk_p}(t-t_p)\right\}\right]^{l_p}.\label{eq_solSigmap}
\end{align}
\end{subequations}
Here $t_q,t_p<t$ are constants of integration determined by the initial condition, as are $l_q,l_p\in\{-1,+1\}$. [The derivation of Eqs.~\ref{eq_explicitSol1}--\ref{eq_solSigma}, which was omitted from Ref.~\cite{theurel2024incompatible}, is included in Appendix~A.] Streamline plots of $(\sigma_{q}(t),\sigma_{p}(t))$ for three different choices of the ratio $k_p/k_q$ are shown in Fig.~\ref{fig_exact_solution_for_Gaussian_PDF}. As can be readily seen, the conjectured CUR (Eq.~\ref{eq_CUR0}), which for the conjugate and mutually-uncorrelated observables $A=q, B=p$ reduces to
\begin{equation}\label{eq_classicalUncertaintyRel1}
	\sigma_q\sigma_p\geq\frac{\qbar}{2},
\end{equation}
indeed holds true in the intended sense: epistemic states that would breach this inequality are unreachable starting from states of low information.

While this example provides an explicit and provocative case in favor of the CUR, it only proves the relation for a narrow class of observables and measurement protocols. Not included in this class are: non-trivial Liouville dynamics ($H\nequiv0$), higher-dimensional phase spaces, measurements of observables other than $q$ and $p$, and time-dependent protocols in which qualities, strengths, efficiencies, or the observables subject to measurement, are explicit functions of time. (Also not included in the example are non-canonical measurements. But as the theory of such measurements has not yet been developed, I do not address them in this paper.) Among the time-dependent measurement protocols, one might further distinguish \emph{open-loop} protocols, in which $\qbar(t),k(t),\nu(t), H(\boldsymbol z, t), A(\boldsymbol z,t)$ are specified ahead of time, and \emph{closed-loop} protocols, in which such time-dependencies are chosen contingent on the stream of measurement outcomes~\cite{fn12}. In attempting to prove the CUR in the general case, it is daunting at this time to proceed by explicit integration of Eq.~\ref{eq_masterEquation}. Instead, my approach will be to seek a local (i.e.,~differential) condition that speaks to the difficulty of advancing into the region of belief space where Eq.~\ref{eq_CUR0} is violated. I draw inspiration for the form of such a condition from Fig.~\ref{fig_exact_solution_for_Gaussian_PDF}: Notice that for states in the inaccessible shaded region, all measurements considered in the example yield net losses of information at each moment in time ($\dot I<0$). This manifests as the streamlines in that region ``pushing out'' against the descending flow, preventing trajectories above the hyperbola from crossing down. In the general case, if one could show that all measurements of quality no greater than $1/\qbar$ must yield net losses of information whenever the epistemic state violates Eq.~\ref{eq_CUR0} for any pair of observables $A,B$, then that would lead partway towards establishing the general validity of the CUR. The next sections present partial results in this direction: A subset of such measurements will be shown to yield net losses of information, in expectation.

\subsubsection*{Two mathematical identities}

\paragraph{A symplectic Cram\'er-Rao bound.}

For any observable $A$ and any PDF $\rho$, consider the function over phase space $\{A,\log\rho\}$. This always has mean of zero, as can be checked using integration by parts:
\begin{align}
	&\langle\{A,\log\rho\}\rangle=\int_M d^{2n}\boldsymbol z\rho\{A,\log\rho\}\\
	&\quad=\int_M d^{2n}\boldsymbol zA\{\log\rho,\rho\}=0,
\end{align}
where the omitted boundary terms vanish due to the factor of $\rho$. Its covariance with any other observable $B$ is
\begin{align}
	&\text{cov}(\{A,\log\rho\},B)\notag\\
	&\quad=\langle\{A,\log\rho\}B\rangle-\langle\{A,\log\rho\}\rangle\langle B\rangle\\
	&\quad=\int_M d^{2n}\boldsymbol z\rho\{A,\log\rho\}B=\int_M d^{2n}\boldsymbol z\{A,\rho\}B\\
	&\quad=\int_M d^{2n}\boldsymbol z\rho\{B,A\}=\langle\{B,A\}\rangle,
\end{align}
where I used the chain rule followed by integration by parts. For any pair of observables $A,B$, consider the $3\times3$ covariance matrix of $\{A,\log\rho\}, A, B$, which is a positive semidefinite matrix:
\begin{align}
	\boldsymbol 0&\preceq\boldsymbol\Sigma_{(\{A,\log\rho\}, A, B)}\\
	&\quad=\left(
	\begin{array}{ccc}
		\sigma_{\{A,\log\rho\}	}^2	&0	&\langle\{B,A\}\rangle\\
		0	&\sigma_A^2	&\text{cov}(A,B)\\
		\langle\{B,A\}\rangle	&\text{cov}(A,B)	&\sigma_B^2
	\end{array}
	\right).
\end{align}
Taking the determinant yields the identity inequality
\begin{equation}\label{eq_symplecticCR1}
	\sigma_{\{A,\log\rho\}}^2(\det\boldsymbol\Sigma_{AB})\geq\sigma_A^2\langle\{A,B\}\rangle^2.
\end{equation}
As far as I know this is the first time this identity has appeared in the literature. A weakening of it,
\begin{equation}\label{eq_symplecticCR2}
	\sigma_{\{A,\log\rho\}}^2\sigma_{B}^2\geq\langle\{A,B\}\rangle^2,
\end{equation}
can be recognized as a ``symplectification'' of a well-known statistical inequality. Indeed, if $(q_1,\dots,q_n,p_1,\dots,p_n)$ is a canonical coordinate system with $q_1=A$, then
\begin{equation}
	\sigma_{\{A,\log\rho\}}^2=\langle\{A,\log\rho\}^2\rangle=\left\langle\left(\frac{\partial\log\rho}{\partial p_1}\right)^2\right\rangle
\end{equation}
is known as the Bayesian Fisher information of $p_1$; so for the special case of $p_1=B$ Eq.~\ref{eq_symplecticCR2} reduces to the Bayesian Cram\'er-Rao bound~\cite{van2001detection}:
\begin{equation}
	\left\langle\left(\frac{\partial\log\rho}{\partial B}\right)^2\right\rangle\sigma_{B}^2\geq1.
\end{equation}

\paragraph{An identity for pairs of affinely related observables.}

Suppose observables $A,B$ are affine functions of $C,D$:
\begin{equation}\label{eq_affineTransf}
	\begin{pmatrix}
		A(\boldsymbol z)\\
		B(\boldsymbol z)
	\end{pmatrix}
	=
	\begin{pmatrix}
		L_{11}	&L_{12}\\
		L_{21}	&L_{22}
	\end{pmatrix}
	\begin{pmatrix}
		C(\boldsymbol z)\\
		D(\boldsymbol z)
	\end{pmatrix}
	+
	\begin{pmatrix}
		c_1\\
		c_2
	\end{pmatrix}.
\end{equation}
Then a straightforward calculation yields $\{A,B\}=(\det\boldsymbol L)\{C,D\}$,
from which follows $\langle\{A,B\}\rangle^2=(\det\boldsymbol L)^2\langle\{C,D\}\rangle^2$. Meanwhile, under the affine transformation the covariance matrix transforms as $\boldsymbol\Sigma_{AB}=\boldsymbol L\boldsymbol\Sigma_{CD}\boldsymbol L^T$, and therefore $\det\boldsymbol\Sigma_{AB}=(\det\boldsymbol L)^2\det\boldsymbol\Sigma_{CD}$. One thus has the identity
\begin{equation}\label{eq_okLemma2}
 	\left(\det\boldsymbol\Sigma_{CD}\right)\left\langle\{A,B\}\right\rangle^2=\left(\det\boldsymbol\Sigma_{AB}\right)\left\langle\{C,D\}\right\rangle^2.
\end{equation}

\subsubsection*{Evidence for the uncertainty relation in the general case}

Starting from Eq.~\ref{eq_antiHtheorem2} one has, for any observable $C$, any epistemic state $\rho$, and any $k\geq0,\nu\in[0,1]$ and $\qbar'\geq\qbar>0$,
\begin{align}
	\underset{(C,\qbar',k,\nu)}{\llangleBig\frac{dI}{dt}\rrangleBig}&=4k\nu \sigma_C^2-k\qbar'^2\sigma_{\{C,\log\rho\}}^2\label{eq_antiHtheorem3}\\
	&\leq4k\left(\sigma_C^2-\left(\frac{\qbar}{2}\right)^2\sigma_{\{C,\log\rho\}}^2\right)\label{eq_antiHtheorem4},
\end{align}
where the underset tuple $(C,\qbar',k,\nu)$ keeps track of the physical situation under consideration: in this case, a canonical measurement of $C$ at quality $1/\qbar'$, strength $k$ and efficiency $\nu$. Let $D$ be any observable. Multiplying through by the non-negative quantity $\det\boldsymbol\Sigma_{CD}$ and applying identity~\ref{eq_symplecticCR1}:
\begin{equation}\label{eq_antiHtheorem5}
	\left(\det\boldsymbol\Sigma_{CD}\right)\underset{(C,\qbar',k,\nu)}{\llangleBig\frac{dI}{dt}\rrangleBig}\leq4k\sigma_C^2\left(\det\boldsymbol\Sigma_{CD}-\left\langle\frac{\qbar}{2}\{C,D\}\right\rangle^2\right).
\end{equation}
This result can be slightly further generalized. Let observables $A,B$ be any affine functions of $C,D$. Multiplying Eq.~\ref{eq_antiHtheorem5} through by the non-negative quantity $\det\boldsymbol\Sigma_{AB}$, dividing through by $\det\boldsymbol\Sigma_{CD}$ (assumed to be non-zero and finite---such edge cases can be included by a continuity argument), and using identity~\ref{eq_okLemma2}, one gets
\begin{equation}\label{eq_antiHtheorem6}
	\left(\det\boldsymbol\Sigma_{AB}\right)\underset{(C,\qbar',k,\nu)}{\llangleBig\frac{dI}{dt}\rrangleBig}\leq4k\sigma_C^2\left(\det\boldsymbol\Sigma_{AB}-\left\langle\frac{\qbar}{2}\{A,B\}\right\rangle^2\right).
\end{equation}
This is the main result. It follows from here that whenever the CUR (Eq.~\ref{eq_CUR0}) is violated for some pair of observables $A, B$ and quantity of action $\qbar$, then any canonical measurement of any affine function of $A,B$ (i.e.,~an observable $C=aA+bB+c$), done at quality no greater than $1/\qbar$, will produce in expectation a net loss of information ($\llangle\dot I\rrangle<0$). In this sense, the CUR controls the acquisition of information at the classical microscale. In particular, this speaks to a certain difficulty of reaching (or remaining in) epistemic states that violate Eq.~\ref{eq_CUR0}; as, intuitively, it seems implausible that a breach in the uncertainty relation could avoid mending for very long while information is constantly being lost. The next section fleshes out this intuition in a special case.

\subsubsection*{Restricted proof of the uncertainty relation for the Heisenberg algebra on systems of one degree of freedom}

On the 2-dimensional phase space $\mathbb R^2$ let $q,p$ be a pair of conjugate observables; for example, but not necessarily, physical position and momentum. Consider the class of observables that are affine functions of $q,p$:
\begin{equation}
	\mathfrak h=\{(q,p)\mapsto aq+bp+c\mid a,b,c\in\mathbb R\}.
\end{equation}
This is known as the Heisenberg algebra; it is a notable Lie subalgebra of the full algebra of observables [which in our classical scenario is $C^\infty(\mathbb R^2)$]. Let observables $A,B,C\in\mathfrak h$, and consider a canonical measurement of $C$ done at quality $1/\qbar'\leq1/\qbar$. There holds Eq.~\ref{eq_antiHtheorem6}, which I abbreviate here as
\begin{gather}\label{eq_ratchet1}
	\llangle\dot I(t)\rrangle\leq x(t)^2\left(1-u(t)\right),
\end{gather}
where $u(t)=\frac{\left\langle\frac{\qbar}{2}\{A,B\}\right\rangle^2}{\det\boldsymbol\Sigma_{AB}}$ and $x(t)^2>0$. Now invoke a well-known identity inequality between information and the covariance matrix,
\begin{equation}\label{eq_informationVarianceInequality}
	-\frac12\log\det\left(2\pi e\boldsymbol\Sigma_{qp}\right)\leq I,
\end{equation}
which expresses the fact that among all PDFs with a given covariance matrix, the Gaussian has the minimal information~\cite{cover1999elements}. Making use of identity~\ref{eq_okLemma2}, and the fact that $\langle\{q,p\}\rangle^2=1$, one can write Eq.~\ref{eq_informationVarianceInequality} as
\begin{gather}\label{eq_ratchet2}
	u(t)\leq a^2\exp\{2I(t)\},
\end{gather}
where $a^2>0$ is a constant. It can now be seen that if, at any time during the course of the measurement of $C$, the epistemic state is such that $u>1$ (i.e., such that the CUR is violated for observables $A,B$), then the two relations~\ref{eq_ratchet1},~\ref{eq_ratchet2} together act as a one-way ratchet that drives $u\to1$. Indeed, by the first relation, $I$ tends to decrease whenever $u>1$; and by the second relation, the decreasing $I$ quickly forces $u$ to decrease too. This ratchet only stops when $u\to1$.

I conclude that, for a system with a single degree of freedom, and with the possible exception of small transient fluctuations, it is impossible, by means of canonical measurements of quality no greater than $1/\qbar$ of observables in the Heisenberg algebra, to reach any epistemic state violating the CUR (Eq.~\ref{eq_CUR0}) for any pair of observables in the Heisenberg algebra. Compared to the example presented in an earlier section (``Illustration of the uncertainty relation in an analytically solvable measurement protocol''), this result extends the proof of the CUR in the following important ways. 
(i) The observable subject to measurement may now be different from the two observables tested in the CUR; each of these observables is now allowed to be any element of the Heisenberg algebra (i.e., any affine combination of $q$ and $p$). (ii) Arbitrary Liouville dynamics ($H\nequiv0$) are now allowed. (iii) Time-dependent protocols are now allowed, with $k(t),\nu(t), H(\boldsymbol z, t)$ completely arbitrary, so long as the (inverse) quality remains bounded $\qbar'(t)\geq\qbar$ and the observable subject to measurement is of the form $C(\boldsymbol z,t)=a(t)q+b(t)p+c(t)$. And (iv) arbitrary initial PDFs $\rho(\boldsymbol z,0)$ are now allowed.

\subsection*{Discussion}

The partial results presented above raise the possibility that the CUR may turn out to be a general theorem belonging to resource theories of classical measurement, expressing a deep epistemic restriction in Hamiltonian mechanics. Such a result would have profound implications, for instance, for the field of classical information thermodynamics~\cite{parrondo2015thermodynamics}, which has developed under the almost universal assumption of over-idealized classical measurements~\cite{fn1}. Yet, such a strong result remains conjectural and further work will be needed to reveal the whole picture. Here I point out what I consider to be the two deepest issues requiring attention. The first is rather glaring: The present study is based on a single model of the measurement process; the canonical model, employing the particular interaction in Eq.~\ref{eq_canonicalModel}. Before any result can be proved in complete generality, there will need to be developed a model-independent theory of classical measurement, analogous to the model-independent theory of quantum measurement~\cite{ozawa1984quantum, nielsen2002quantum}. 

The second issue is more subtle. The theory of measurement in Ref.~\cite{theurel2024incompatible}, on which the present study is based, relies on the assumption of ``designer Hamiltonians'' (DHs); taking for granted that any desired interaction (such as Eq.~\ref{eq_canonicalModel}) can be instantiated exactly and for free. As it turns out, this assumption seems to enable violations of the CUR when filtering or post-selecting from large ensembles of similar systems~\cite{theurel2024incompatible}. However, I suspect this assumption to be an over-idealization that will need to be revised in a more careful analysis: The instantiation of a Hamiltonian should not be treated as a theoretical primitive, but as a theory-laden problem dependent on one's control capabilities; the better these are, the more precisely one will be able to instantiate a target Hamiltonian---but also, I suspect, the higher the thermodynamic cost will be, and certain Hamiltonians may be altogether impossible to instantiate. The latter notion is suggested by the Wigner-Araki-Yanase theorem~\cite{loveridge2011measurement}, which proves, in the quantum setting, that the presence of globally-conserved additive observables constrains the interaction Hamiltonians that can be exactly instantiated. I'm not aware of any classical analogue of this theorem having been developed, but I expect it to be possible. Regarding the thermodynamic cost of instantiating a target Hamiltonian, support for this notion may be drawn from the thermodynamic uncertainty relations (TURs)~\cite{barato2015thermodynamic, horowitz2020thermodynamic}. Indeed, consider an interaction Hamiltonian that has free parameters appearing in it; for example the canonical interaction, Eq.~\ref{eq_canonicalModel}, corresponding to a measurement of the observable
\begin{equation}
	A(q,p)=\frac12\left((q-q_0)^2+(p-p_0)^2\right).
\end{equation}
[It is interactions of this form, dependent on perfectly known free parameters, that seem to enable violations of the CUR in Ref.~\cite{theurel2024incompatible}.] The free parameters $q_0,p_0$ are treated by the assumption of DHs as given constants. But they must actually be controlled degrees of freedom of a larger physical system; and the TURs associate a minimal thermodynamic cost to the precise control of certain degrees of freedom. These remarks all point to finite-temperature \emph{Hamiltonian control} as a topic deserving renewed consideration, with the potential to reveal significant amendments to the present results and to those in Ref.~\cite{theurel2024incompatible}. It is in the resource theories belonging to that amended theory of measurement that I conjecture the CUR to be a general epistemic restriction.


\clearpage 

%
\bibliography{limits_of_knowledge} 

\begin{thebibliography}{10}
\bibitem{fn0}
The expression ``fundamental theory'' is used here in the philosophy-of-science
  sense; as in contrast to narrower theories within the special sciences.
\bibitem{fn2}
A further reason for such lapse must surely be the monumental rise of quantum
  theory in the early twentieth century. The fact that quantum mechanics has a
  theory of measurement built into it, which proved to be extraordinarily
  successful at describing precision experiments, relieved the pressure that
  would otherwise have continued to mount to examine the measurement process in
  classical terms. (This is not to suggest that all aspects of real-world
  precision experiments could be satisfyingly understood in purely classical
  terms.).
\bibitem{fn6}
All stochastic differential equations in this article are to be interpreted in
  the It\^o sense.
\bibitem{fn7}
Different agents, with different states of knowledge, reading the same outcome
  $A^*$, will break it down differently between ``signal,'' $\langle A\rangle$,
  and ``noise,'' $\frac{1}{\sqrt{8k}}\frac{dW}{dt}$.
\bibitem{fn10}
Here expectation values are taken with respect to any quantum state $\hat\rho$;
  $\sigma_{\hat A},\sigma_{\hat B}$ are defined in their usual
  quantum-mechanical way, and $\text{cov}(\hat A,\hat B)=\langle\hat A\hat
  B+\hat B\hat A\rangle/2-\langle\hat A\rangle\langle\hat B\rangle$.
\bibitem{fn12}
The quantities $\qbar,k,\nu$ as well as the observables $H,A$ appearing in
  Eq.~\ref{eq_masterEquation} may all be explicit functions of time and the
  equation remains valid without modification. But caution must be taken if
  converting from It\^o to the equivalent Stratonovich SDE, as the resulting
  equation will have terms involving the variations $\delta k/\delta \rho$,
  $\delta \nu/\delta \rho$ and $\delta A/\delta\rho$, and so the Stratonovich
  SDE will be more complicated for closed-loop than for open-loop protocols.
\end{thebibliography}
\bibliographystyle{sciencemag}

%
%
%
%
%
%


\section*{Acknowledgments}
I thank O. Eulogio L\'opez for reading a version of the typescript and providing valuable feedback.

\paragraph*{Competing interests:}
There are no competing interests to declare.


\newpage


\appendix

\subsection*{Appendix A.\hspace{1em}Analytical solution of a simple measurement protocol}

My objective is to analytically solve Eq.~\ref{eq_masterEquation} for the following measurement protocol. On the 2-dimensional phase space $\mathbb R^2=\{(q,p)\}$, the simultaneous continuous canonical measurement of $q$ and $p$, at fixed quality $1/\qbar$, fixed respective strengths $k_q,k_p$, and perfect efficiencies $\nu_q=\nu_p=1$, in the absence of internal dynamics ($H\equiv0$). The version of Eq.~\ref{eq_masterEquation} adapted to simultaneous measurements of multiple observables is derived in Ref.~\cite{theurel2024incompatible}. In this case it is
\begin{align}
	\frac{\partial\rho}{\partial t}&=k_q\qbar^2\frac{\partial^2\rho}{\partial p^2}+k_p\qbar^2\frac{\partial^2\rho}{\partial q^2}+\sqrt{8k_q}(q-\langle q\rangle)\rho\frac{dW_q}{dt}\notag\\
	&\quad+\sqrt{8k_p}(p-\langle p\rangle)\rho\frac{dW_p}{dt},\label{eq_masterEquation2}
\end{align}
where $W_q, W_p$ are independent Wiener processes. This is a nonlinear It\^o stochastic PDE in the unknown $\rho(q,p,t)$. I seek to solve it subject to the initial condition
\begin{equation}\label{eq_initCond}
	\rho(q,p,0)=\frac{1}{2\pi\sigma_{q}(0)\sigma_{p}(0)}e^{-\frac{(q-\mu_{q}(0))^2}{2{\sigma_{q}(0)}^2}-\frac{(p-\mu_{p}(0))^2}{2{\sigma_{p}(0)}^2}},
\end{equation}
which is an uncorrelated gaussian in $q$ and $p$. To this end, I try an ansatz solution that preserves this form over time:
\begin{equation}\label{eq_ansatz}
	\rho(q,p,t)=\frac{1}{2\pi\sigma_{q}(t)\sigma_{p}(t)}e^{-\frac{(q-\mu_{q}(t))^2}{2{\sigma_{q}(t)}^2}-\frac{(p-\mu_{p}(t))^2}{2{\sigma_{p}(t)}^2}},
\end{equation}
where $\mu_q(t),\mu_p(t),\sigma_q(t),\sigma_p(t)$ are yet-to-be-determined functions of time, which may be stochastic. Working on the left-hand side (l.h.s.) of Eq.~\ref{eq_masterEquation2}, I substitute in the ansatz:
\begin{align}
	\text{(l.h.s.)}dt&=\rho(q,p,t+dt)-\rho(q,p,t)\\
	&=\frac{e^{-\frac{(q-\mu_{q}-d\mu_{q})^2}{2(\sigma_{q}+d\sigma_{q})^2}-\frac{(p-\mu_{p}-d\mu_{p})^2}{2(\sigma_{p}+d\sigma_{p})^2}}}{2\pi(\sigma_{q}+d\sigma_{q})(\sigma_{p}+d\sigma_{p})}-\rho(q,p,t),
\end{align}
and Taylor-expand to second order in the small quantities $d\mu_q,d\mu_p,d\sigma_q,d\sigma_p$. After some tedious algebra I get:
\begin{align}
	\text{(l.h.s.)}dt&=\rho\bigg(
	-\frac{d\sigma_q}{\sigma_q}
	-\frac{d\sigma_p}{\sigma_p}
	+\frac{d\mu_q}{\sigma_q}\bar q
	+\frac{d\mu_p}{\sigma_p}\bar p
	+\frac{d\sigma_q}{\sigma_q}\bar q^2
	\notag\\
	&\qquad
	+\frac{d\sigma_p}{\sigma_p}\bar p^2
	+\frac{d\sigma_q^2}{\sigma_q^2}
	+\frac{d\sigma_p^2}{\sigma_p^2}
	+\frac{d\sigma_q}{\sigma_q}\frac{d\sigma_p}{\sigma_p}
	\notag\\
	&\qquad
	-3\frac{d\mu_q}{\sigma_q}\frac{d\sigma_q}{\sigma_q}\bar q
	-\frac12\frac{{d\mu_q}^2}{\sigma_q^2}
	-\frac52\frac{{d\sigma_q}^2}{\sigma_q^2}\bar q^2
	\notag\\
	&\qquad
	+\frac12\frac{{d\mu_q}^2}{\sigma_q^2}\bar q^2
	+\frac12\frac{{d\sigma_q}^2}{\sigma_q^2}\bar q^4
	+\frac{d\mu_q}{\sigma_q}\frac{d\sigma_q}{\sigma_q}\bar q^3
	\notag\\
	&\qquad
	-3\frac{d\mu_p}{\sigma_p}\frac{d\sigma_p}{\sigma_p}\bar p
	-\frac12\frac{{d\mu_p}^2}{\sigma_p^2}
	-\frac52\frac{{d\sigma_p}^2}{\sigma_p^2}\bar p^2
	\notag\\
	&\qquad
	+\frac12\frac{{d\mu_p}^2}{\sigma_p^2}\bar p^2
	+\frac12\frac{{d\sigma_p}^2}{\sigma_p^2}\bar p^4
	+\bar p^3\frac{d\mu_p}{\sigma_p}\frac{d\sigma_p}{\sigma_p}
	\notag\\
	&\qquad
	-\frac{d\mu_p}{\sigma_p}\frac{d\sigma_q}{\sigma_q}\bar p
	-\frac{d\sigma_q}{\sigma_q}\frac{d\sigma_p}{\sigma_p}\bar p^2
	-\frac{d\mu_q}{\sigma_q}\frac{d\sigma_p}{\sigma_p}\bar q
	\notag\\
	&\qquad
	-\frac{d\sigma_q}{\sigma_q}\frac{d\sigma_p}{\sigma_p}\bar q^2
	+\frac{d\mu_q}{\sigma_q}\frac{d\mu_p}{\sigma_p}\bar q\bar p
	+\frac{d\mu_q}{\sigma_q}\frac{d\sigma_p}{\sigma_p}\bar q\bar p^2
	\notag\\
	&\qquad
	+\frac{d\mu_p}{\sigma_p}\frac{d\sigma_q}{\sigma_q}\bar q^2\bar p
	+\frac{d\sigma_q}{\sigma_q}\frac{d\sigma_p}{\sigma_p}\bar q^2\bar p^2
	\bigg),\label{eq_lhs}
\end{align}
where I've abbreviated $\bar q=\frac{q-\mu_q}{\sigma_q}, \bar p=\frac{p-\mu_p}{\sigma_p}$. Meanwhile, substituting the ansatz~\ref{eq_ansatz} into the right-hand side (r.h.s.) of Eq.~\ref{eq_masterEquation2} yields
\begin{align}
	\text{(r.h.s.)}dt&=
	\rho\bigg(
	-\frac{k_q\qbar^2dt}{\sigma_p^2}
	-\frac{k_p\qbar^2dt}{\sigma_q^2}
	+\frac{k_q\qbar^2dt}{\sigma_p^2}\bar p^2
	+\frac{k_p\qbar^2dt}{\sigma_q^2}\bar q^2\notag\\
	&\qquad
	+\sqrt{8k_q}\sigma_qdW_q\bar q
	+\sqrt{8k_p}\sigma_pdW_p\bar p
	\bigg).\label{eq_rhs}
\end{align}
Equating coefficients of like powers of $\bar q$ and $\bar p$ between Eqs.~\ref{eq_lhs},~\ref{eq_rhs} yields a set of conditions which must hold to first order in $dt$. The conditions corresponding to powers of degrees three and four are
\begin{gather}
		{d\sigma_q}^2={d\sigma_p}^2=0,\label{eq_aux3}\\
	d\mu_qd\sigma_q=d\mu_pd\sigma_p=d\mu_qd\sigma_p=d\mu_pd\sigma_q=d\sigma_qd\sigma_p=0.\label{eq_aux4}
\end{gather}
The condition corresponding to $\bar q^1\bar p^1$ is
\begin{equation}
		d\mu_qd\mu_p=0.\label{eq_aux5}
\end{equation}
Using Eq.~\ref{eq_aux4}, the conditions corresponding to $\bar q^1$ and $\bar p^1$ simplify to
\begin{gather}
	d\mu_q
	=
	\sqrt{8k_q}\sigma_q^2dW_q,\label{eq_aux6}
	\\
	d\mu_p
	=
	\sqrt{8k_p}\sigma_p^2dW_p.\label{eq_aux7}
\end{gather}
Using Eqs.~\ref{eq_aux3}--\ref{eq_aux7}, the conditions corresponding to $\bar q^2$ and $\bar p^2$ simplify to
\begin{gather}
	\frac{1}{\sigma_q}\frac{d\sigma_q}{dt}
	+4k_q\sigma_q^2
	=
	\frac{k_p\qbar^2}{\sigma_q^2},\label{eq_aux8}
	\\
	\frac{1}{\sigma_p}\frac{d\sigma_p}{dt}
	+4k_p\sigma_p^2
	=
	\frac{k_q\qbar^2}{\sigma_p^2}.\label{eq_aux9}
\end{gather}
Finally, using Eqs.~\ref{eq_aux3}--\ref{eq_aux7}, the condition corresponding to $\bar q^0\bar p^0$ simplifies to
\begin{equation}\label{eq_aux10}
	\frac{1}{\sigma_q}\frac{d\sigma_q}{dt}
	+\frac{1}{\sigma_p}\frac{d\sigma_p}{dt}
	+4k_q\sigma_q^2
	+4k_p\sigma_p^2
	=
	\frac{k_p\qbar^2}{\sigma_q^2}
	+\frac{k_q\qbar^2}{\sigma_p^2},
\end{equation}
which is already implied by Eqs.~\ref{eq_aux8},~\ref{eq_aux9}, and so contributes nothing new. In deriving Eqs.~\ref{eq_aux8}--\ref{eq_aux10} I made use of the rule of It\^o calculus:
\begin{equation}
	dW_q^2=dW_p^2=dt.\label{eq_aux1}
\end{equation}
In the next step I will also make use of the complementary rules:
\begin{equation}
	dt^2=dW_qdt=dW_pdt=dW_qdW_p=0.\label{eq_aux2}
\end{equation}
Eqs.~\ref{eq_aux3}--\ref{eq_aux9} constitute a set of necessary and sufficient conditions for the solution: If there exist functions $\mu_q(t),\mu_p(t),\sigma_q(t),\sigma_p(t)$ whose increments satisfy all of these conditions, then the ansatz~\ref{eq_ansatz} will be the unique solution to the initial-value problem~\ref{eq_masterEquation2},~\ref{eq_initCond}. Now, the most general form for the increments is
\begin{align}
	d\mu_q&=x_q(t)dt+y_q(t)dW_q+z_q(t)dW_p,\label{eq_aux11}\\
	d\mu_p&=x_p(t)dt+y_p(t)dW_q+z_p(t)dW_p,\label{eq_aux12}\\
	d\sigma_q&=\xi_q(t)dt+\lambda_q(t)dW_q+\zeta_q(t)dW_p,\label{eq_aux13}\\
	d\sigma_p&=\xi_p(t)dt+\lambda_p(t)dW_q+\zeta_p(t)dW_p,\label{eq_aux14}
\end{align}
where the 12 coefficients appearing here are unknown functions of time. Making use of Eqs.~\ref{eq_aux1},~\ref{eq_aux2}, condition~\ref{eq_aux3} is equivalent to $\lambda_q^2+\zeta_q^2=\lambda_p^2+\zeta_p^2\equiv0$, or
\begin{equation}\label{eq_aux15}
	\lambda_q=\zeta_q=\lambda_p=\zeta_p\equiv0.
\end{equation}
So $\sigma_q(t),\sigma_p(t)$ must be deterministic functions. This also ensures that conditions~\ref{eq_aux4} are all satisfied. Condition~\ref{eq_aux5} is automatically satisfied in view of Eqs.~\ref{eq_aux6},~\ref{eq_aux7}. The latter equations give $\mu_q(t),\mu_p(t)$ reduced to quadratures in terms of $\sigma_q(t),\sigma_p(t)$. So all that remains is to find deterministic functions $\sigma_q(t),\sigma_p(t)$ satisfying the final set of conditions~\ref{eq_aux8},~\ref{eq_aux9}. Fortunately, the latter are well-posed deterministic ODEs for $\sigma_q(t),\sigma_p(t)$. It can be verified by direct substitution that their general solution is
\begin{align}
	\sigma_{q}(t)^2&=\frac{\qbar}{2}\sqrt{\frac{k_p}{k_q}}\left[\coth\left\{4\qbar\sqrt{k_qk_p}(t-t_q)\right\}\right]^{l_q},\label{eq_aux16}\\
	\sigma_{p}(t)^2&=\frac{\qbar}{2}\sqrt{\frac{k_q}{k_p}}\left[\coth\left\{4\qbar\sqrt{k_qk_p}(t-t_p)\right\}\right]^{l_p},\label{eq_aux17}
\end{align}
where $t_q,t_p<t$ are constants of integration determined by the initial condition, as are $l_q,l_p\in\{-1,+1\}$. With this, all conditions have been met and the problem is solved.

\end{document}